\documentclass[aps,twocolumn,showpacs,prl]{revtex4}
\usepackage[dvips]{graphicx}
\usepackage[english]{babel}
\usepackage[T1]{fontenc}
\usepackage{mathrsfs}
\usepackage[tbtags]{amsmath}
\usepackage{amssymb}
\usepackage{amsxtra}
\usepackage{amsopn}
\usepackage{latexsym}
\usepackage[mathcal]{eucal}
\usepackage{mathtools}

\newcommand{\BE}{\begin{equation}}
\newcommand{\EE}{\end{equation}}
\newcommand{\BA}{\begin{align}}
\newcommand{\EA}{\end{align}}
\newcommand{\Tr}{\mathrm Tr}
\newcommand{\nn}{\nonumber}

\newcommand{\ppp}{ \frac{{\rm d}^4p}{(2\pi)^4}}
\newcommand{\fsl}[1]{\ensuremath{\mathrlap{\!\not{\phantom{#1}}}#1}}
\begin{document}

\title{The Principle of Stationary Variance in Quantum Field Theory}

\author{Fabio Siringo}

\affiliation{Dipartimento di Fisica e Astronomia 
dell'Universit\`a di Catania,\\ 
INFN Sezione di Catania,
Via S.Sofia 64, I-95123 Catania, Italy}

\date{\today}
\begin{abstract}
The principle of stationary variance is advocated as a viable variational approach
to quantum field theory. The method is based on the principle that the variance
of energy should be at its minimum when the state of a quantum system reaches its best 
approximation for an eigenstate. While not too much popular in quantum mechanics, the method
is shown to be valuable in quantum field theory, and three special examples are given in very
different areas ranging from Heisenberg model of antiferromagnetism to quantum electrodynamics
and gauge theories.
\end{abstract}


\maketitle

Since when Lord Rayleigh described his method for calculating the frequencies
of a mechanical system in 1873\cite{rayleigh}, the variational method
has become very popular. In quantum mechanics (QM), the variational method follows from the well known
property that the expectation value of the Hamiltonian is stationary when the quantum
state is an eigenstate. However, that property is shared with any function of the Hamiltonian,
and a more general variational method can be introduced by looking for the stationary points
of other functions of the Hamiltonian\cite{sigma,gep2}. Among them, the variance has the important
physical features of being positive, bounded from below, and vanishing at the exact eigenstates.
Moreover, by Heisenberg relations, its finite value gives a measure of the life-time of an
approximate eigenstate. 

\begin{figure}[b] \label{fig:graphs}
\centering
\includegraphics[width=0.45\textwidth,angle=-90]{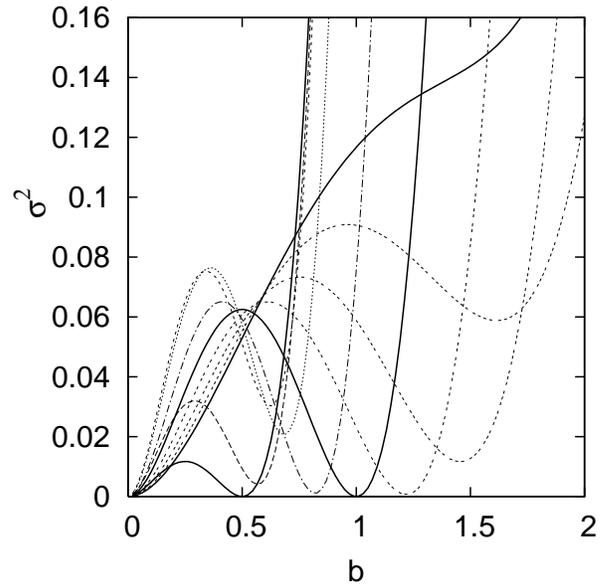}
\caption{The variance for the trial state of the hydrogen atom Eq.(\ref{hyd}), for
$\eta=$ 1.0 (solid line), 0.8, 0.6, 0.4, 0.2, 0.0 (solid line), -0.2, -0.4, -0.6, -0.8 (solid line). 
The minimum moves
from $b=0.5$ at $\eta=1$ (exact excited state) to $b=1$ at $\eta=0$ (exact ground state),
and then raises for $\eta<0$ as the trial state progressively worsen, 
disappearing for $\eta<-0.7$.}
\end{figure}

While the principle of stationary variance is not very popular in QM,
it can be very useful in quantum field theory (QFT). In this letter we show that
in QFT the search for the stationary points of the variance provides a
viable variational approach to selected problems where the standard variational
method is known to fail or give trivial results. Should a strong coupling preclude
the use of perturbation theory, the stationary variance would be a valid alternative
to numerical lattice simulations. We give three examples of well known extended physical systems
where the standard variational method fails to produce an acceptable description, 
while the principle of stationary
variance provides reasonable nontrivial results: the Heisenberg limit of the half-filled Hubbard model
of antiferromagnetism, the simple theory of a self-interacting scalar field, and quantum
electrodynamics (QED).

In QM, denoting by $\langle {\cal O}\rangle$ the expectation value of the operator
${\cal O}$ in the state $\vert \Psi\rangle$, the variance of the Hamiltonian $H$ 
can be written as
$\sigma^2=\langle H^2\rangle-\langle H\rangle^2$ and
satisfies
\BE
{\langle\Psi\vert\Psi \rangle}\frac{\delta \sigma^2}{\delta \langle\Psi \vert}=
\left(H^2-\langle H^2\rangle\right)\vert \Psi\rangle-2\langle H\rangle\left(
H-\langle H\rangle\right) \vert \Psi\rangle.
\EE
Thus the first variation of $\sigma^2$ is zero if the state $\vert \Psi\rangle$
is an eigenstate of $H$. Since $\sigma^2\ge 0$ in any state, and $\sigma^2=0$ in eigenstates,
whenever a trial state approaches an eigenstate the variance is expected to
be stationary and to show a local minimum. Of course that happens for any eigenstate,
not just the ground state, and some caution is required when the trial state can
approach different eigenstates. In Ref.\cite{sigma} a detailed discussion of the method
is reported for simple problems of QM. For instance in the simple case of an hydrogen atom,
and a trial state 
\BE
\langle r\vert\Psi\rangle=(1-\eta b r) e^{-b r}
\label{hyd}
\EE
that, in atomic units, is the exact ground state for $\eta=0, b=1$ and the first excited state
for $\eta=1, b=0.5$, the variance is reported in Fig.1 as a function of $b$ for several values of $\eta$.
We observe a pronounced minimum when the trial state approaches one of the
eigenstates.

In QM, the predictive power of the standard variational method can be improved by just increasing
the number of free parameters, thus enlarging the subspace spanned by the trial wave function.
There is no real utility in a more complex second order calculation, like that of variance, 
that would require the matrix elements of the square of $H$.
In QFT, because of calculability, the trial functional must be Gaussian, the standard
variational method leads to the Gaussian effective potential (GEP) 
\cite{schiff,rosen,barnes,stevenson,ibanez,var,light,bubble,superc1,superc2,abreu,kim,su2,LR,AF,HT}
and there is no obvious way to improve the approximation. In fact, several extended
physical systems are described by field theories that are not suited to be described by
a first order approximation like the GEP. Second order terms might give important
contributions that could be captured by a second order variational
method like that of stationary variance.
 
As a first example, let us consider the half-filled two-dimensional Hubbard model of a narrow band
conductor with a strong on-site repulsive correlation\cite{fradkin}
\BE
H=-t\sum_{<ij>,\alpha}C^\dagger_{i\alpha} C_{j\alpha}+U\sum_i n_{i+}n_{i-}
\EE
where $C^\dagger_{i\alpha}$ ($C_{i\alpha})$ are creation (annihilation) operators for the electrons, 
$\alpha=\pm\>$ is the spin projection, $n_{i\alpha}=C^\dagger_{i\alpha}C_{i\alpha}$ are number operators
and the site indices $i,j$ run over first neighbors on a square lattice. The Fermi liquid is known to
be unstable towards an antiferromagnetic (AF) ground state\cite{fradkin}. If $U$ is large, we can take 
the simultaneous eigenstates of the number operators as a basis set, and each of these states 
can be labeled by a string of charges $n_{i\alpha}=0,1$, with $\sum_{i\alpha} n_{i\alpha}=N$ where
$N$ is the total number of electrons that is assumed to be equal to the number of lattice sites.
For $t=0$ there is a massive degeneracy in the system: the ground state has an energy $E=0$ and
is given by any linear combination of the degenerate $2^N$ states $\vert m\rangle_0$
with no double occupancy and a single electron on each site (with $n_{i+}+n_{i-}=1$); the first 
excited state has $E=U$ and is given by the degenerate states $\vert m\rangle_1$
with a single double occupancy (and a single hole), etc. 
We expect that the degeneracy would be removed by a small but finite hopping term
$t\ll U$. In this strong coupling limit, it would be reasonable to take as a trial
ground state the linear combination $\vert\Psi\rangle=\sum_m a_m \vert m\rangle_0$
while the hopping term can be regarded as a small perturbation. It is quite obvious that
first order perturbation theory and standard variational method are useless in the
present case: if we define by $H_0=H_{t=0}$ the correlation term, and by $V_t=H_{U=0}$
the hopping term, so that $H=H_0+V_t$,
even for $t\not= 0$ we always find 
\BE
_0\langle m^\prime \vert V_t\vert m\rangle_0=0
\label{Vt}
\EE
and then $\langle \Psi\vert H\vert \Psi\rangle =0$.
The excited states cannot be neglected even when they are very far at $E=U$, so that we need
the second order perturbative correction or a second order variational method if we want to keep
the trial state $\vert \Psi\rangle$ in the ground state subspace. In fact the failure of the first
order approximations is due to the vanishing of all first order matrix elements in the ground state
subspace according to Eq.(\ref{Vt}).

Let us look at the variance: since  $\langle \Psi\vert H\vert\Psi\rangle=0$, the
variance can be written as
\BE
\sigma^2=\langle \Psi\vert H^2\vert\Psi\rangle
=\sum_m {\langle \Psi \vert V_t\vert m\rangle_1} \cdot{_1\langle m\vert V_t\vert\Psi\rangle}
\EE
where the sum runs over first excited states only, and 
$\vert_1\langle m\vert V_t \vert m^\prime\rangle_0\vert^2=t^2$
when the states differ for the hopping of a single electron between first neighbor sites or
vanishes otherwise. In fact, these matrix elements allow for some electron motion among first neighbors,
and the matrix elements of the variance $_0\langle m\vert H^2 \vert m^\prime\rangle_0$ differ from
zero in the ground state subspace, if the states $m\not=m^\prime$ differ by a spin flip of two
first neighbor electrons. Actually, because of Pauli principle,the matrix elements of $H^2$ enumerate all 
the first neighbor pairs of electrons with opposite spin that can hop and go back
from one site to the
other  by the second order process. By a straightforward calculation, in the single occupancy
subspace, $H^2$ can be written up to a constant as an effective Hamiltonian\cite{fradkin}
in terms of local spin operators $\vec S_i$
\BE
H^2=-2t^2\sum_{<ij>} \vec S_i\cdot \vec S_j=-U H_H
\EE
where $H_H$ is the well known Heisenberg Hamiltonian of the spin system.
We conclude that, in the single occupancy subspace spanned by $\vert\Psi\rangle$,
the eigenstates of the variance are the eigenstates of the Heisenberg Hamiltonian that is
known to be the exact strong coupling limit of the Hubbard model at half-filling,
and has an AF ground state.
Thus, in the same subspace, the stationary states of the variance must be the eigenstates of the
Heisenberg model, and the principle of the stationary variance yields the correct
AF ground state of the system. 
This simple example tells us that the same method of the stationary variance could be useful 
for gauge theories where the minimal coupling does not give any contribution to the effective 
potential in first order approximations like the GEP. We need a Lagrangian version of the method
that has been developed for the simple theory of a self-interacting scalar field\cite{sigma,gep2}.

Actually, even for the simple scalar theory, the standard variational description by the GEP presents
some shortcomings that can only be cured by a second order approximation.
For instance the order of the transition that is known to be second order, but is weakly first
order in the GEP\cite{stevenson}. The problem is solved by inclusion of second order terms in a 
post Gaussian effective potential (PGEP)\cite{stancu}, but the second order effective potential is
not bounded from below, and has no stationary points. On the other hand the method of stationary variance
is perfectly viable, and yields a second order variational approximation that improves the GEP and
predicts a second order transition\cite{sigma}. Thus the scalar theory is the best example for
illustrating the Lagrangian approach to the method. The Lagrangian reads
 \BE
{\cal L}=\frac{1}{2}\partial^\mu\phi\>\partial_\mu\phi
-\frac{1}{2} m_B^2\phi^2-\lambda\phi^4.
\label{Lbare}
\EE
We define a shifted field
$h=\phi-\varphi$ where $\varphi$ is a constant background, and
split the action functional as $S[\phi]=S_0[h]+S_I[h]$ where
$S_0[h]$ is an arbitrary trial Gaussian functional, quadratic in the fields, that can be thought as
the action of a free particle theory. It can be written as
\BE
S_0[h]=\frac{1}{2}\int h(x) g^{-1} (x,y) h(y) {\rm d}^4x{\rm d}^4y
\EE 
where $g(x,y)$ is an unknown trial propagator.
One of the main merits of the Lagrangian approach is that
the effective action $\Gamma[\varphi]$ can be evaluated as a sum of 
Feynman diagrams by the general representation\cite{weinbergII}
\BE
e^{i\Gamma[\varphi]}= \int_{1PI} {\cal D}_h
e^{iS[\varphi+h]}
=\int_{1PI} {\cal D}_h e^{iS_0[h]} e^{iS_I[h]} 
\label{path}
\EE
equivalent to the sum of all the one-particle-irreducible (1PI) vacuum diagrams for
the action functional $S[\varphi+h]$, with $S_I$ that plays the role of the interaction.
In terms of the quantum average
\BE
\langle {\cal O}\rangle = \frac{ \int_{1PI} {\cal D}_h e^{iS_0[h]} {\cal O}}
{ \int {\cal D}_h e^{iS_0[h]} },
\EE
the effective action can be written as an expansion in moments of $S_I$
\begin{align}
i\Gamma_n[\varphi]=\sum_{n=0}^{\infty}  i\Gamma_n[\varphi]&=i\Gamma_0+
\langle iS_I \rangle+\frac{1}{2!} 
\langle [ iS_I-\langle iS_I\rangle]^2 \rangle\nn\\
&+\frac{1}{3!} \langle [ iS_I-\langle iS_I\rangle]^3\rangle+\dots
\label{exp}
\end{align}
that is the sum of all connected  1PI vacuum diagrams, while $\Gamma_0$
follows exactly from the quadratic $S_0$. 
Thus we can use perturbation theory for evaluating the effective potential
$V(\varphi)=-\Gamma[\varphi]/\Omega$ where $\Omega$ is the total space-time volume.
On the other hand, since the exact action does not depend
on the arbitrary choice of $S_0$ (and $S_I$), we can optimize the splitting of $S$ by
a variational criterion that makes
the effects of the interaction $S_I$ smaller in the vacuum of $S_0$, yielding
a convergent expansion even without any small parameter\cite{minimal}.
The principle of stationary variance suggests itself, since by Eq.(\ref{exp}) we
see that the second order term of the effective potential is 
$V_2=-\sigma_I^2/{2\Omega}$
where $\sigma_I$ is the variance of the Euclidean form of $S_I$, as
follows immediately by Wick rotating.
The variance would be zero if the vacuum of $S_0$ were an exact eigenstate of $S_I$, while a
minimal variance is expected to optimize the convergence of the expansion. It is quite obvious
that $\sigma_I$ is equal to the variance of the total action $S$, because powers of $S_0$ only give
disconnected contributions. Thus in this approach the variance of the Lagrangian is used
instead of the variance of the Hamiltonian.
The free parameters can be fixed by a stationary condition for the second order term of the
effective potential $V_2$, yielding a second order variational criterion. In fact, while
the variance is bounded and $\sigma_I^2>0$, the second order effective potential is not, as the
minimum of $V_2$ would be a maximum for $\sigma^2_I$. Actually, by insertion of a free particle
trial propagator $g^{-1}(k)=k^2-M^2$ the stationary (minimum) point of $\sigma_I$ yields a solution
for the mass $M$, and the corresponding second order potential shows a continuous phase transition
improving on the simple first order GEP\cite{sigma}. However, the approximation can be improved 
further by considering any functional form for $g^{-1}(k)$, and imposing the stationary condition
by the functional condition $\delta V_2/\delta g=0$ that becomes an integral equation for the
optimal propagator $g$. Moreover, there is no need to evaluate the effective potential as the
stationary condition can be derived by the self-energy directly, making use of 
the general connection
\BE
\frac{\delta V_n}{\delta g(k)}=\frac{i}{2} \left( \Sigma_n (k)-\Sigma_{n-1}(k)\right)
\label{delVns}
\EE
where $\Sigma_n$ is the nth-order self-energy term. 
This connection follows by Wick's theorem, and a detailed derivation will be published elsewhere\cite{gep2}.
Taking $n=2$, the integral equation for the trial
propagator $g$ can be simply written as $\Sigma_2=\Sigma_1$.

This machinery seems to be suited for theories with gauge interacting fermions, 
since the minimal coupling has no effect
on the first order potential, and other variational approaches like GEP and PGEP give 
trivial results\cite{stancu2}. That is a circumstance that we already encountered in the
Heisenberg limit of the Hubbard model. Let us consider the simple case of QED 
with a single massive fermion 
\BE
{\cal L}=\bar\Psi(i\fsl{\partial}+e  {\ensuremath{\mathrlap{\>\not{\phantom{A}}}A}}
-m)\Psi-\frac{1}{4}F^{\mu\nu}F_{\mu\nu}
-\frac{1}{2}(\partial_\mu A^\mu)^2    
\label{Le}
\EE
where the last term is the gauge fixing term in Feynman gauge, and $F_{\mu\nu}$ is
the electromagnetic tensor.
We must introduce two trial propagators in $S_0$, one $D_{\mu\nu}(k)$ for photons, 
and the other $G^{ab}(k)$ for fermions. These are unknown trial functions, 
and in general differ from 
the bare propagators
$\Delta_{\mu\nu}^{-1} (k)=-\eta_{\mu\nu} k^2$, $g_m^{-1}(k)=\fsl{k}-m$.

The stationary conditions for the effective potential can be derived by the
general connection to self-energy and polarization functions
\BE
\frac{\delta V_n}{\delta G^{ab} (k)}=-i \left( \Sigma^{ba}_n (k)-\Sigma^{ba}_{n-1}(k)\right).
\label{delVnS}
\EE
\BE
\frac{\delta V_n}{\delta D_{\mu\nu} (k)}=\frac{i}{2} \left( \Pi^{\nu\mu}_n (k)-\Pi^{\nu\mu}_{n-1}(k)\right)
\label{delVnP}
\EE
These equations generalize Eq.(\ref{delVns}), and their detailed derivation will be given in Ref.\cite{gep2}.
The criterion of stationary variance is enforced by imposing that
$\delta V_2/\delta G=0$ and $\delta V_2/\delta D=0$, that according to Eqs.(\ref{delVnS}),
(\ref{delVnP}) are equivalent to $\Sigma_2=\Sigma_1$ and $\Pi_2=\Pi_1$. Self-energy and polarization 
functions follow by use of the standard perturbation theory with an optimized interaction $S_I=S-S_0$,
as for the scalar theory. While the standard variational method gives the trivial result
$D=\Delta$, $G=g_m$, the stationary equations for the variance can be written as
\begin{align}
G(k)&=g_m(k)-g_m(k)\cdot \Sigma_2^\star (k)\cdot g_m (k)\nn\\
D_{\mu\nu}(k)&=\Delta_{\mu\nu}(k)-\Delta_{\mu\lambda}(k)\cdot 
{\Pi_2^\star}^{\lambda\rho}(k)\cdot \Delta_{\rho\nu}(k)
\label{stationary}
\end{align}
where $\Pi_2^\star$, $\Sigma_2^\star$ are the usual proper two-point functions
\begin{align}
\Sigma_2^\star (k)&=i e^2 \int \ppp \gamma^\mu G(k+p)\gamma^\nu D_{\mu\nu}(p) \nn\\
{\Pi_2^\star} ^{\mu\nu}(k)&=-i e^2 \int \ppp
\Tr\left\{ G(p+k) \gamma^\mu  G(p)\gamma^\nu\right\}. 
\label{proper}
\end{align}
Of course these one-loop terms contain divergences, and a regularization scheme must
be adopted. That is not a difficult task, as in the present Lagrangian approach we can
use the standard techniques of perturbation theory\cite{stancu}, and renormalize bare 
parameters and functions order by order. A detailed description of renormalization
by dimensional regularization will be given in Ref.\cite{varqed}. 

The stationary equations Eqs.(\ref{stationary}) are a set of coupled integral
equations, and their numerical solution would be equivalent to the sum of an
infinite set of Feynman graphs. They are expected to hold even in the limit of 
strong coupling, as they derive from a variational criterion, and lead
to nontrivial physical insights. 
While a numerical analysis
is out of the aim of the present paper, we anticipate that, by a spectral
representation, and with some constraint, the numerical problem can be recast in a
set of coupled Volterra integral equations that have a unique solution and can be
solved by iterative methods\cite{varqed}.

The functions $G$, $D$ can be regarded as the building blocks of an optimized expansion,
and it is instructive to study their higher order corrections. For instance,
the second order propagator $G^{(2)}$ can be written in terms
of the proper self-energy as
\BE
G^{(2)} (k)=\left[ G^{-1}(k)-\Sigma_1(k)-\Sigma_2^\star(k)\right]^{-1}
\EE
and by an explicit calculation
\BE
[G^{(2)} (k)]^{-1}=\fsl{k}-m -\Sigma_2^\star(k)
\label{G2b}
\EE
that resembles the one-loop result of QED, but with the functions $G$ and $D$
substituted  
in the one-loop $\Sigma_2^\star$, in Eq.(\ref{proper}), instead of the bare
propagators $g_m$, $\Delta$. Expanding the stationary conditions Eqs.(\ref{stationary}) in
powers of the coupling $e^2$, taking the lowest order approximation $G\approx g_m$, 
$D\approx\Delta$, and substituting back in the proper self energy $\Sigma_2^\star$, then
Eq.(\ref{G2b}) would become exactly equal to the one-loop propagator of QED. 
Thus we conclude that in the weak coupling limit the principle of stationary variance
reproduces the standard results of QED.

While the strong coupling limit of QED is not of any real phenomenological interest, the stationary
variance could be an important tool for a non-perturbative analytical study of non-Abelian gauge theories
and QCD, whenever the large strength of the interaction does not
allow the use of perturbation theory. Since we have shown that
the principle of stationary variance provides reasonable results in very different areas of physics, 
we expect that its extension to non-Abelian gauge theories might be very useful 
for a better understanding of the low energy
phenomenology of strong interactions.

\end{document}